\documentclass[12pt,preprint]{aastex}

\shorttitle{Non-uniform Be-$\alpha$ relationships for the Galaxy}
\shortauthors{Tan \& Zhao}

\begin{document}

\title{A possible signature of non-uniform Be-$\alpha$ relationships
       for the Galaxy\altaffilmark{\star}}

\author{Kefeng Tan\altaffilmark{1} and Gang Zhao\altaffilmark{1}}

\altaffiltext{$\star$}{Based in part on data obtained from the
    ESO/ST-ECF Science Archive Facility.}
\altaffiltext{1}{Key Laboratory of Optical Astronomy, National
    Astronomical Observatories, Chinese Academy of Sciences, Beijing
    100012, China; [tan, gzhao]@nao.cas.cn}

\begin{abstract}
Most of the previous studies on beryllium abundances in metal-poor
stars have taken different Galactic populations as a whole when
investigating the production and evolution of Be. In this Letter, we
report on the detection of systematic differences in
[$\alpha$/H]-$A$(Be) relationships between the low- and high-$\alpha$
stars which were identified by previous works. We remind that one
should be more careful in investigating the Galactic evolution of Be
with a sample comprising different Galactic populations, because
such a mixed sample may lead to inaccurate Be-Fe/Be-O relationships.
\end{abstract}


\section{Introduction}\label{intro}
During the past two decades, astronomers have been working on
beryllium (Be) abundances in metal-poor stars to investigate the
production and evolution of Be in the early Galaxy. Most of these
studies have taken different Galactic populations (halo and thick-disk
stars, if included) as a whole when investigating the Be-Fe and Be-O
relationships. This may be partly due to that the production of Be is
thought to be a global process across the Galaxy, and therefore Be
abundances should show small scatter at a given metallicity
(\citealt{suzuki2001}), no matter which population the stars belong
to. \citet{boesgaard2006} presented, for the first time, some
evidences of Be abundances dispersions. \citet{smiljanic2008} and
\citet{tan2009} also found some Be-rich stars which are obviously
above the general Be-Fe and Be-O trends. In their Be abundance survey
with the largest sample to date, \citet[hereafter S09]{smiljanic2009}
found a marginal statistical detection of a real scatter in the
[Fe/H]-$A$(Be)\footnote{$\mathrm{[A/B]}=\log[N(\mathrm{A})/N(\mathrm{B}
)]_{\bigstar}-\log[N(\mathrm{A})/N(\mathrm{B})]_{\odot}$, $A(\mathrm{X})
=\log[N(\mathrm{X})/N(\mathrm{H})]+12$} diagram. They proposed two
possible explanations for this: one is that the halo and the
thick-disk stars have different [Fe/H]-$A$(Be) relationships; the
other is that Be abundances dispersions exist at any given
metallicity.

It is now generally believed that the Galaxy was formed through
hierarchical merging, which means that different components of the
Galaxy may have experienced different chemical evolution histories.
Since $\alpha$-elements are mainly produced by core collapse
supernovae, they are closely associated with the star formation
history of the Galaxy. In this regard, Galactic components with
different $\alpha$-elements abundances patterns may show different
behaviors in Be abundances. \citet{nissen1997} found evidence of a
bimodal distribution of [$\alpha$/Fe] ($\alpha$ refers to the average
abundance of Mg, Si, Ca, and Ti) for about a dozen halo stars.
\citet{gratton2003a,gratton2003b} found that the accretion component
of the Galaxy tends to have lower [$\alpha$/Fe] than the dissipative
component. S09 also claimed two distinct components of the Galactic
halo, but in the log(Be/H)-[$\alpha$/Fe] diagram, which seems to make
the division clearer than in the [Fe/H]-[$\alpha$/Fe] diagram.
Recently, \citet[hereafter NS10]{nissen2010} performed a precise
abundance analysis for a sample of 94 thick-disk and halo stars in the
solar neighborhood. They confirmed that, in the [Fe/H]-[$\alpha$/Fe]
diagram, the halo stars separate into two distinct populations, i.e.
the low- and high-$\alpha$ halo populations. The very high precision
of stellar parameters and $\alpha$-elements abundances make the sample
of NS10 ideal for studying the possible different Be abundance
patterns for different Galactic components in a systematic way. In
this Letter, we report on the detection of two systematically
different Be-$\alpha$ relationships for the stars from the sample of
NS10.

\section{The sample}\label{sample}
Among the 94 stars in the sample of NS10, high-resolution and high
signal-to-noise ratio VLT/UVES spectra covering the \ion{Be}{2}
3130\,{\AA} resonance doublet were available from the ESO/ST-ECF
Science Archive for 40 stars. We downloaded the reduced spectra for
these stars and then determined their Be abundances using the same
technique as described in \citet{tan2009}.
In addition, another 3 stars have Be abundances from the literature (2 from
\citealt{rich2009} and 1 from \citealt{boesgaard1993}). For these 3
stars, in order to keep our analysis consistent, we first derived
their equivalent widths of the \ion{Be}{2} 3130\,{\AA} resonance lines
based on the Be abundances and stellar parameters from the literature,
and then re-calculated their Be abundances. For all the sample stars, stellar
parameters given by NS10 were adopted in the determination of Be
abundances. The uncertainty of Be abundance mainly comes from the
uncertainties of stellar parameters and pseudo-continuum location.
NS10 estimated the uncertainties of effective temperature, surface
gravity, and metallicity to be $\pm$30\,K, $\pm$0.05\,dex, and
$\pm$0.04\,dex, respectively, which leads to a total uncertainty of
about $\pm$0.03\,dex in Be abundance. The typical uncertainty of Be
abundance introduced by pseudo-continuum location was estimated to be
$\pm$0.05\,dex. In total, the typical internal uncertainty of Be
abundance is around $\pm$0.06\,dex. Abundances and uncertainties for
the $\alpha$-elements were taken from NS10 directly. The abundances
results for the sample stars are given in Table~\ref{table:par}.
We noted that, among the 40 stars whose Be abundances were determined
directly from the VLT/UVES spectra, 33 stars were included in the
sample of S09. If the Be abundances given by S09 are scaled to the
stellar parameters of NS10, then the average difference between our
results and that of S09 will be $0.03\pm0.07$\,dex (ours minus S09's),
which is in reasonable agreement within uncertainties.

In summary, we have consistent Be and $\alpha$-elements abundances for
43 stars, of which 14 are thick-disk stars, 13 are low-$\alpha$ halo
stars, and 16 are high-$\alpha$ halo stars according to NS10
(low-/high-$\alpha$ means low/high [$\alpha$/Fe] at a given [Fe/H]). As
mentioned in NS10, the classification of thick-disk and halo is based
on kinematic criterion which is somehow uncertain (if the velocity
distribution of the thick-disk is non-Gaussian with an extended tail
toward high velocities, then some of the high-$\alpha$ halo stars
might belong to the thick-disk population). Therefore, in this work we
classify the sample stars into two categories only, i.e. the low- and
high-$\alpha$ stars, based purely on their $\alpha$-elements abundances.
In other words, the low-$\alpha$ stars in this work correspond to the
low-$\alpha$ halo stars in NS10, and the high-$\alpha$ stars in this
work include both the thick-disk and the high-$\alpha$ halo stars in
NS10. The reasonableness of classifying the thick-disk and the
high-$\alpha$ halo stars into the same category can be further
supported by the indistinguishable abundance patterns of more elements
for these two populations as shown in NS10 and \citet{nissen2011}.
\citet{gratton2003a,gratton2003b} also mentioned that no clear
discontinuity between the rotating part of the halo (mostly
high-$\alpha$ halo stars) and the thick-disk could be found.

\section{Results and discussion}\label{results}
\begin{figure}
\centering
\includegraphics{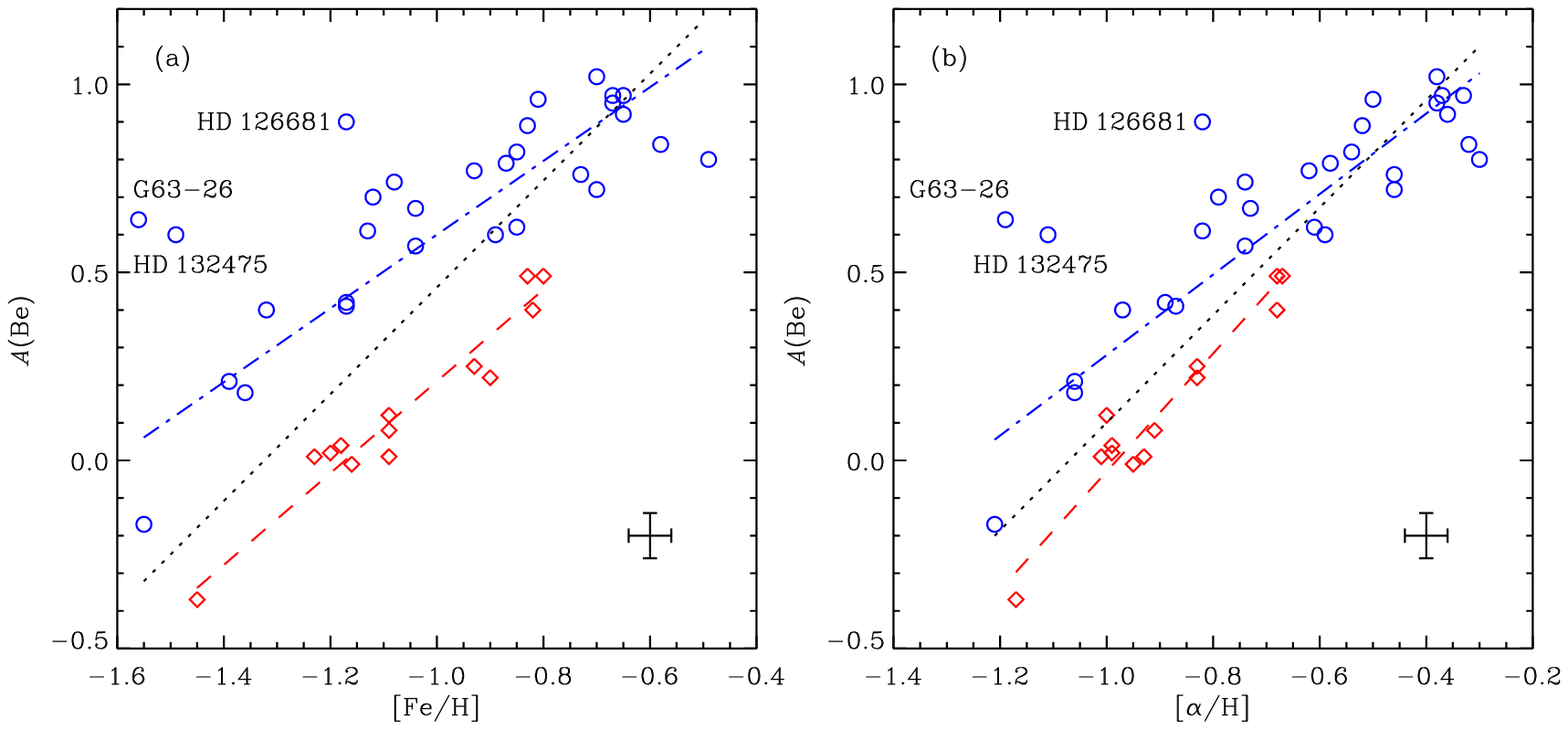}
\caption{Be abundance as a function of (a) [Fe/H] and (b) [$\alpha$/H]
for the sample stars. Circles are high-$\alpha$ stars and diamonds are
low-$\alpha$ stars. Dash-dotted, dashed, and dotted lines are the best
linear fits for high-$\alpha$ stars, low-$\alpha$ stars, and the whole
sample, respectively.\label{fig:be_alpha}}
\end{figure}

\begin{figure}
\centering
\includegraphics{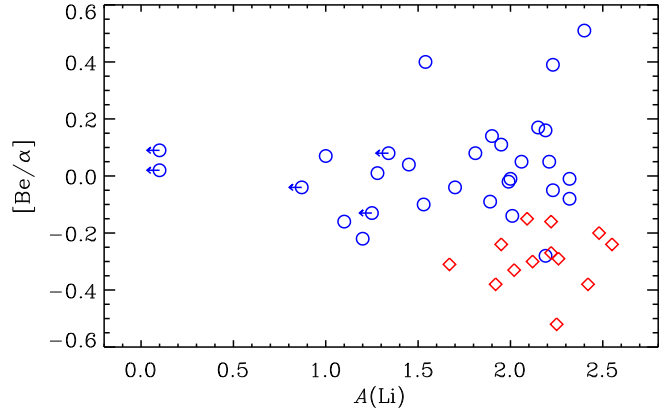}
\caption{[Be/$\alpha$] as a function of $A$(Li) for the sample stars.
The symbols have the same meanings as in Figure~\ref{fig:be_alpha};
the symbols with leftward arrows represent the upper limits of Li
abundances.\label{fig:li}}
\end{figure}

\begin{figure}
\centering
\includegraphics{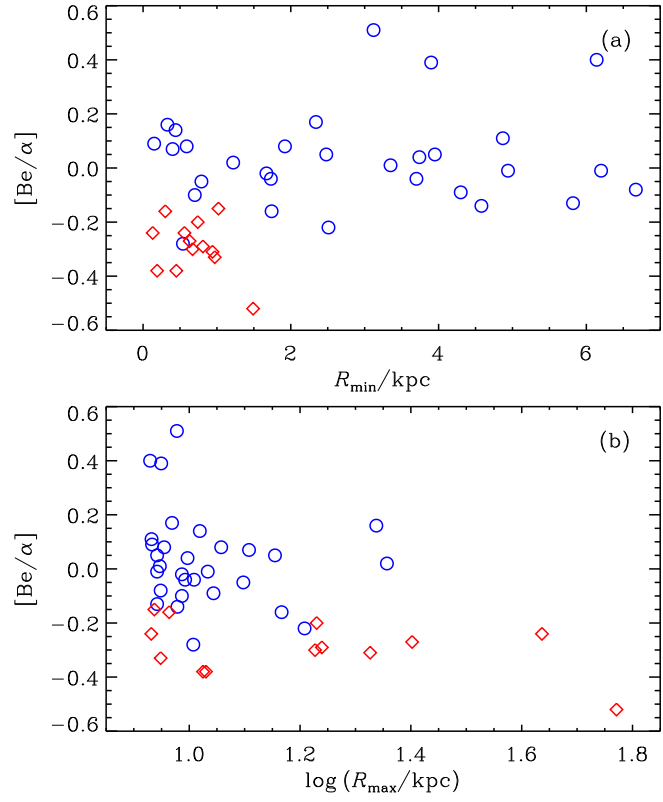}
\caption{[Be/$\alpha$] as a function of (a) perigalactic and (b)
apogalactic distances for the sample stars. The symbols have the same
meanings as in Figure~\ref{fig:be_alpha}.\label{fig:kin}}
\end{figure}

Figure~\ref{fig:be_alpha} shows $A$(Be) as a function of [Fe/H] and of
[$\alpha$/H] for the sample stars. Here we use [$\alpha$/H] as a proxy
for [O/H] as it is generally believed that $\alpha$-elements have the
same origin as oxygen. Similar approximation has also been adopted by
S09. It can be seen clearly in Figure~\ref{fig:be_alpha}(a) that Be
abundances of the low-$\alpha$ stars are systematically lower than
that of the high-$\alpha$ stars, and the differences are obviously
larger than the uncertainties of Be abundances. Even so, one may still
expect a uniform $A$(Be) versus [$\alpha$/H] relationship for the
sample stars because the production of Be is thought to be  correlated
directly with oxygen (and thus $\alpha$-elements) rather than iron.
However, as can be seen in Figure~\ref{fig:be_alpha}(b), Be abundances
of the low-$\alpha$ stars are still systematically lower than that of
the high-$\alpha$ stars. As the differences in Be abundances in the
[$\alpha$/H]-$A$(Be) diagram are not as obvious as that in the
[Fe/H]-$A$(Be) diagram, we made a statistical test to confirm whether
the division is a real feature or just caused by the uncertainties of
the analysis. We made linear fits, in the common [$\alpha$/H] range
($-1.2<[\alpha/\mathrm{H}]<-0.6$), to the high-$\alpha$ stars, to the
low-$\alpha$ stars, and to the whole sample\footnote{All through this
Letter, G63-26, HD\,132475, and HD\,126681 are not included in any
linear fits; please refer to the last paragraph but one of this Letter
for more discussions about these three stars.}, and then calculated
the root mean square (RMS) deviations from the fits respectively. The
RMS deviations from the fits for the high- and low-$\alpha$ stars are
0.07 and 0.06\,dex, respectively, which are comparable to the Be
abundances uncertainty (0.06\,dex); while the RMS deviation for the
whole sample is 0.18\,dex, which is obviously larger than the error of
Be abundances. Therefore, we conclude that systematic differences in
[$\alpha$/H]-$A$(Be) relationships do exist between the low- and
high-$\alpha$ stars in our sample. One may think that Be in the
low-$\alpha$ stars have been depleted and thus the abundances are
lower than that of the high-$\alpha$ stars. This is, however, not the
case. As shown in Figure~\ref{fig:li}, Li abundances of the
low-$\alpha$ stars are in the upper range of the sample, but their
[Be/$\alpha$] values (solar Be abundance $A(\mathrm{Be})=1.32$
recommended by \citealt{lodders2010} was adopted in the calculation)
are systematically lower than that of the high-$\alpha$ stars. As the
destruction temperature for Li is lower than that of Be, if Be is
depleted, then Li should have been depleted more than Be. The
possibility that the gas from which the low-$\alpha$ stars formed had
been exposed to the cosmic-rays shorter than that of the high-$\alpha$
stars and thus has lower Be abundances can also be excluded, because
the low-$\alpha$ stars are on average 2--3\,Gyr younger than the
high-$\alpha$ stars (Schuster et al. 2011, in preparation).
\citet{pasquini2005} proposed the possibility that some stars may form
at very large Galactocentric radii with lower cosmic-ray fluxes and
thus lower production of Be and heavy elements. However, as shown in
Figure~\ref{fig:kin}, we didn't find any gradient in Be abundances for
the sample stars. In the Be abundance survey with the largest sample
to date by S09, no decreasing trend in Be abundance with increasing
Galactocentric distance was found, either.

S09 identified 4 pairs of stars with similar atmospheric parameters
and metallicities but different Be abundances to testify the existence
of scatter in Be abundances. We suggest that one of the main reasons
for the scatter is the composition of the pairs. Among these 4 pairs
of stars, 3 pairs are included in our sample, and each pair is
comprised of one low-$\alpha$ star and one high-$\alpha$ star.
According to the definition of NS10, at a given [Fe/H], low-$\alpha$
stars have lower [$\alpha$/Fe] (and thus [$\alpha$/H]) than
high-$\alpha$ stars. Moreover, at a given [$\alpha$/H], low
[$\alpha$/Fe] stars have less Be than high [$\alpha$/Fe] stars as
shown in Figure~\ref{fig:be_alpha}(b). Therefore, it is natural that
low-$\alpha$ stars have lower Be abundances than high-$\alpha$ stars
with the same metallicities. Also, it can be easily understood that
the separation between the low- and high-$\alpha$ stars in the
[Fe/H]-$A$(Be) diagram is clearer than that in the [$\alpha$/H]-$A$(Be)
diagram (compare the two panels of Figure~\ref{fig:be_alpha}). S09
noted the impression of two parallel relations in the Be-Fe diagram.
However, they did not detect the different [$\alpha$/H]-$A$(Be)
relationships. This may be partly due to the relatively large
abundance uncertainties of S09 (note that the stellar parameters and
$\alpha$-elements abundances adopted by S09 were taken from different
works and had not been homogenized).

As suggested by NS10, the high-$\alpha$ stars may form in regions with
rapid chemical evolution, while the low-$\alpha$ stars may originate
from dwarf galaxies with lower star formation rates. In this case, the
energy spectrum for the cosmic-rays may be different in the regions
where the low- and high-$\alpha$ stars formed, and thus the production
rate of Be could be different (as the effective spallation
cross-sections are dependent on the energies of the cosmic-rays).
Therefore, systematic differences in Be abundances between the low-
and high-$\alpha$ stars can be a natural prediction from the
explanation of NS10. We note that the bimodal distribution of
[$\alpha$/Fe] observed by NS10 is consistent with the theoretical
results of \citet{zolotov2009,zolotov2010}, which are based on the
assumption that the inner halos of the galaxies contain both stars
accreted from satellite galaxies and stars formed in situ. However,
\citet{nissen2011} shows that abundance patterns of the low-$\alpha$
stars do not match that of the present-day dwarf galaxies. They
suggest that the low-$\alpha$ stars may have been accreted from more
massive satellite galaxies at early times.

As shown in Figure~\ref{fig:be_alpha}(b), in the common [$\alpha$/H]
range ($-1.2<[\alpha/\mathrm{H}]<-0.6$), $A$(Be) of the high-$\alpha$
stars increases with [$\alpha$/H] faster than that of the low-$\alpha$
stars; however, if the ``metal-rich'' end of the high-$\alpha$ stars
are included, then the [$\alpha$/H]-$A$(Be) relationship for the
high-$\alpha$ stars will be flatter than that for the low-$\alpha$
stars. Such a break in the [$\alpha$/H]-$A$(Be) relationship makes it
look like that the ``metal-rich'' end of the high-$\alpha$ stars are
following the relation defined by the low-$\alpha$ stars. Similar
change in slope also seems to appear in the [Fe/H]-$A$(Be)
relationship at $\mathrm{[Fe/H]}\sim-1$ as can be seen in
Figure~\ref{fig:be_alpha}(a). Inspired by the prediction of shallower
slope at higher metallicity from the mass outflow models,
\citet{boesgaard1999} investigated the possible change in slope in the
Be-Fe relationship. But the metallicity range covered by this work is
relatively narrow, and our sample is a bit small, hence it is
premature to make any conclusions. According to NS10, for halo stars
with $\mathrm{[Fe/H]}<-1.4$ ($[\alpha/\mathrm{H}]<-1.1$), the low- and
high-$\alpha$ populations will merge together in the
[Fe/H]-[$\alpha$/Fe] diagram due to the same origins of iron and
$\alpha$-elements. However, these two populations could still be
distinguished in the [$\alpha$/H]-$A$(Be) diagram only if the
environments where these two populations formed had different energy
spectrum for the cosmic-rays. In this regard, the [$\alpha$/H]-$A$(Be)
diagram can be a useful tool to investigate the properties of the
environments where the stars formed.

We note that there are three stars standing out the general
Be-$\alpha$ (and Be-Fe) trend obviously. HD\,132475 was first found to
be Be-rich by \citet{boesgaard2006} and was later confirmed by
\citet{tan2009}. HD\,126681 was found to be Be-rich by \citet{tan2009}.
G63-26 is a new Be-rich star found in this work; its Be abundance is
about 0.8\,dex higher than that of normal stars with similar
[$\alpha$/H]. According to the results of NS10 and \citet{nissen2011},
G63-26 shows normal Na, Mg, Ca, Ti, Cr, Mn, and Cu abundances, but is
overabundant in Si and Ni. The Y and Ba abundances of this star are
also in the upper range of the sample. The Li abundance of G63-26 is
determined to be $A(\mathrm{Li})=2.4$, which is a bit higher than the
Spite plateau. Such a peculiar abundance pattern is very similar to
that of HD\,106038, which shows obvious overabundances of Li, Si, Ni,
Y, and Ba as described in \citet{smiljanic2008}, though the amplitudes
of abundance deviations from normal stars are much smaller than that
of HD\,106038. This indicates that the gas from which G63-26 formed
may have experienced the same but less intensive chemical enrichment
process as HD\,106038 (for example, polluted by nucleosynthetic
products of Hypernovae as suggested by \citealt{smiljanic2008}).

In summary, we detect systematic differences in [$\alpha$/H]-$A$(Be)
relationships between the low- and high-$\alpha$ stars in the Galaxy.
The [$\alpha$/H]-$A$(Be) diagram for different Galactic populations
can help us understand better the formation and chemical evolution of
the Galaxy. We remind that, using a sample comprising different
Galactic populations to investigate the Galactic evolution of Be may
lead to inaccurate Be-Fe/Be-O relationships. For example, in the whole
metallicity range covered by our sample stars, the equations of linear
fits to the [$\alpha$/H]-$A$(Be) relationships for the high-$\alpha$
stars, for the low-$\alpha$ stars, and for the whole sample are
$A(\mathrm{Be})=(1.07\pm0.06)[\alpha/\mathrm{H}]+(1.35\pm0.04)$,
$A(\mathrm{Be})=(1.57\pm0.17)[\alpha/\mathrm{H}]+(1.54\pm0.15)$, and
$A(\mathrm{Be})=(1.43\pm0.05)[\alpha/\mathrm{H}]+(1.53\pm0.04)$,
respectively (the corresponding equations of the [Fe/H]-$A$(Be)
relationships are
$A(\mathrm{Be})=(0.98\pm0.05)\mathrm{[Fe/H]}+(1.58\pm0.05)$,
$A(\mathrm{Be})=(1.22\pm0.12)\mathrm{[Fe/H]}+(1.43\pm0.13)$, and
$A(\mathrm{Be})=(1.42\pm0.06)\mathrm{[Fe/H]}+(1.88\pm0.06)$,
respectively), which are not the same within uncertainties. However,
we should keep in mind that the sample of this work is relatively
small, and the results need to be confirmed with a larger sample.
Besides, we use the $\alpha$-elements abundances as a proxy for oxygen
abundances, while it is better to investigate the Be-O relationships
directly, though O abundances derived from different indicators are
usually not consistent. Nevertheless, we hope that the results
presented in this work could stimulate more research on Galactic
evolution of Be and related fields, both observational and theoretical.

\acknowledgments
We are grateful to the anonymous referee for the valuable suggestions
and comments. This work is supported by the National Nature Science
Foundation of China under grant Nos.~10821061 and 11003002, by the
National Basic Research Program of China under grant No.~2007CB815103,
and by the Young Researcher Grant of National Astronomical
Observatories, Chinese Academy of Sciences. This research has made use
of the SIMBAD database, operated at CDS, Strasbourg, France and NASA's
Astrophysics Data System.

\clearpage

\clearpage

\begin{deluxetable}{lccrcrcrrcc}
\tabletypesize{\small}
\tablewidth{0pt}
\tablecolumns{11}
\tablecaption{Abundances results, orbital parameters, and classifications for the sample stars\label{table:par}}
\tablehead{\colhead{Star} & \colhead{[Fe/H]} & \colhead{[$\alpha$/H]} & \colhead{$A$(Be)} & \colhead{Ref.} & \colhead{$A$(Li)} & \colhead{Ref.} & \colhead{$R_{\mathrm{min}}$} & \colhead{$R_{\mathrm{max}}$} & \colhead{Ref.} & \colhead{Class\tablenotemark{a}}}
\startdata
BD\,$-21\degr 3420$ & $-1.13$ &  $-0.82$ &   0.61 &(1) &      1.95 &(2) &4.87&  8.55 &(2)&  high-$\alpha$     \\
CD\,$-33\degr 3337$ & $-1.36$ &  $-1.06$ &   0.18 &(1) &      2.32 &(2) &6.67&  8.88 &(2)&  high-$\alpha$     \\
CD\,$-57\degr 1633$ & $-0.90$ &  $-0.83$ &   0.22 &(1) &      2.22 &(2) &0.63& 25.25 &(2)&  low-$\alpha$      \\
CD\,$-61\degr 282$  & $-1.23$ &  $-1.01$ &   0.01 &(1) &      2.12 &(2) &0.67& 16.87 &(2)&  low-$\alpha$      \\
G05-19              & $-1.18$ &  $-0.99$ &   0.04 &(1) &      2.26 &(2) &0.81& 17.35 &(2)&  low-$\alpha$      \\
G05-40              & $-0.81$ &  $-0.50$ &   0.96 &(1) &      1.90 &(2) &0.44& 10.45 &(2)&  high-$\alpha$     \\
G18-28              & $-0.83$ &  $-0.52$ &   0.89 &(1) &$\leq0.10$ &(2) &0.15&  8.57 &(2)&  high-$\alpha$     \\
G18-39              & $-1.39$ &  $-1.06$ &   0.21 &(1) &      2.23 &(6) &0.79& 12.52 &(1)&  high-$\alpha$     \\
G21-22              & $-1.09$ &  $-1.00$ &   0.12 &(3) &      2.48 &(7) &0.74& 16.98 &(1)&  low-$\alpha$      \\
G46-31              & $-0.83$ &  $-0.68$ &   0.49 &(1) &      2.09 &(2) &1.02&  8.65 &(2)&  low-$\alpha$      \\
G63-26              & $-1.56$ &  $-1.19$ &   0.64 &(1) &      2.40 &(1) &3.12&  9.50 &(1)&  (high-$\alpha$)   \\
G121-12             & $-0.93$ &  $-0.83$ &   0.25 &(1) &      2.55 &(2) &0.13& 43.31 &(2)&  low-$\alpha$      \\
G159-50             & $-0.93$ &  $-0.62$ &   0.77 &(1) &      1.00 &(2) &0.40& 12.82 &(2)&  high-$\alpha$     \\
G188-22             & $-1.32$ &  $-0.97$ &   0.40 &(1) &      2.21 &(8) &2.48& 14.28 &(1)&  high-$\alpha$     \\
HD\,3567            & $-1.16$ &  $-0.95$ &$-0.01$ &(1) &      2.42 &(2) &0.19& 10.72 &(2)&  low-$\alpha$      \\
HD\,17820           & $-0.67$ &  $-0.38$ &   0.95 &(1) &      1.28 &(2) &3.35&  8.85 &(2)&  high-$\alpha$     \\
HD\,22879           & $-0.85$ &  $-0.54$ &   0.82 &(1) &      1.45 &(2) &3.74&  9.93 &(2)&  high-$\alpha$     \\
HD\,25704           & $-0.85$ &  $-0.61$ &   0.62 &(1) &      1.89 &(2) &4.30& 11.06 &(2)&  high-$\alpha$     \\
HD\,51754           & $-0.58$ &  $-0.32$ &   0.84 &(1) &      1.10 &(2) &1.74& 14.68 &(2)&  high-$\alpha$     \\
HD\,76932           & $-0.87$ &  $-0.58$ &   0.79 &(1) &      2.06 &(2) &3.95&  8.75 &(2)&  high-$\alpha$     \\
HD\,97320           & $-1.17$ &  $-0.89$ &   0.42 &(1) &      2.32 &(4) &6.20& 10.80 &(4)&  high-$\alpha$     \\
HD\,103723          & $-0.80$ &  $-0.67$ &   0.49 &(1) &      2.22 &(4) &0.30&  9.20 &(4)&  low-$\alpha$      \\
HD\,105004          & $-0.82$ &  $-0.68$ &   0.40 &(1) &      1.95 &(2) &0.56&  8.54 &(2)&  low-$\alpha$      \\
HD\,111980          & $-1.08$ &  $-0.74$ &   0.74 &(1) &      2.19 &(2) &0.33& 21.76 &(2)&  high-$\alpha$     \\
HD\,113679          & $-0.65$ &  $-0.33$ &   0.97 &(1) &      1.99 &(2) &1.67&  9.70 &(2)&  high-$\alpha$     \\
HD\,114762A         & $-0.70$ &  $-0.46$ &   0.72 &(1) &      2.01 &(2) &4.58&  9.52 &(2)&  high-$\alpha$     \\
HD\,120559          & $-0.89$ &  $-0.59$ &   0.60 &(1) &$\leq1.25$ &(2) &5.82&  8.75 &(2)&  high-$\alpha$     \\
HD\,121004          & $-0.70$ &  $-0.38$ &   1.02 &(1) &$\leq1.34$ &(2) &0.59&  9.01 &(2)&  high-$\alpha$     \\
HD\,126681          & $-1.17$ &  $-0.82$ &   0.90 &(1) &      1.54 &(2) &6.14&  8.50 &(2)&  high-$\alpha$     \\
HD\,132475          & $-1.49$ &  $-1.11$ &   0.60 &(1) &      2.23 &(4) &3.90&  8.90 &(4)&  (high-$\alpha$)   \\
HD\,148816          & $-0.73$ &  $-0.46$ &   0.76 &(1) &      1.53 &(2) &0.70&  9.70 &(2)&  high-$\alpha$     \\
HD\,160693          & $-0.49$ &  $-0.30$ &   0.80 &(5) &      1.20 &(5) &2.51& 16.15 &(1)&  high-$\alpha$     \\
HD\,163810          & $-1.20$ &  $-0.99$ &   0.02 &(1) &      1.67 &(2) &0.94& 21.21 &(2)&  low-$\alpha$      \\
HD\,175179          & $-0.65$ &  $-0.36$ &   0.92 &(1) &$\leq0.87$ &(4) &1.73&  9.83 &(2)&  high-$\alpha$     \\
HD\,179626          & $-1.04$ &  $-0.73$ &   0.67 &(1) &      1.81 &(2) &1.92& 11.42 &(2)&  high-$\alpha$     \\
HD\,189558          & $-1.12$ &  $-0.79$ &   0.70 &(1) &      2.15 &(2) &2.34&  9.31 &(2)&  high-$\alpha$     \\
HD\,193901          & $-1.09$ &  $-0.93$ &   0.01 &(1) &      1.92 &(2) &0.45& 10.59 &(2)&  low-$\alpha$      \\
HD\,194598          & $-1.09$ &  $-0.91$ &   0.08 &(1) &      2.02 &(2) &0.97&  8.88 &(2)&  low-$\alpha$      \\
HD\,199289          & $-1.04$ &  $-0.74$ &   0.57 &(1) &      2.00 &(2) &4.94&  8.75 &(2)&  high-$\alpha$     \\
HD\,205650          & $-1.17$ &  $-0.87$ &   0.41 &(1) &      1.70 &(4) &3.70& 10.20 &(4)&  high-$\alpha$     \\
HD\,219617          & $-1.45$ &  $-1.17$ &$-0.37$ &(1) &      2.25 &(2) &1.49& 59.03 &(2)&  (low-$\alpha$)    \\
HD\,222766          & $-0.67$ &  $-0.37$ &   0.97 &(1) &$\leq0.10$ &(2) &1.22& 22.74 &(2)&  high-$\alpha$     \\
HD\,233511          & $-1.55$ &  $-1.21$ &$-0.17$ &(3) &      2.19 &(8) &0.54& 10.17 &(1)&  (high-$\alpha$)   \\
\enddata
\tablerefs{(1) determined by this work; (2) S09; (3) \citet{rich2009}; (4) \citet{tan2009};
(5) \citet{boesgaard1993}; (6) \citet{charbonnel2005}; (7) \citet{boesgaard2006}; (8) \citet{shi2007}.}
\tablenotetext{a}{Parentheses means the classification is uncertain for stars with $\mathrm{[Fe/H]}<-1.4$
according to NS10.}
\end{deluxetable}

\end{document}